\documentclass[12pt,amsmath,amssymb]{revtex4}
\usepackage{amsmath,amsfonts,amsthm,amscd,amssymb,latexsym}
\usepackage{bm}
\usepackage{dcolumn}
\usepackage{graphicx}
\usepackage{epstopdf}
\usepackage{color}
\usepackage{epsf}
\usepackage{epsfig}
\usepackage{graphicx, epic, eepic, color}

\begin{document}

\title{Quasinormal Modes of Generalized Black Holes: \\
$\delta$-Kerr Spacetime}

\author{Alireza \surname{Allahyari}$^{1}$}
\email{alireza.al@ipm.ir}
\author{Hassan \surname{Firouzjahi}$^{1}$}
\email{firouz@ipm.ir}
\author{Bahram \surname{Mashhoon}$^{1,2}$}
\email{mashhoonb@missouri.edu}

\affiliation{$^1$School of Astronomy, Institute for Research in Fundamental
Sciences (IPM), P. O. Box 19395-5531, Tehran, Iran\\
$^2$Department of Physics and Astronomy, University of Missouri, Columbia, Missouri 65211, USA\\
}

\date{\today}

\begin{abstract}
The nonlinear superposition of the $\delta$-metric and the Kerr metric results in $\delta$-Kerr metric that represents a deformed Kerr black hole with $\delta = 1 + q$, where $q > 0$ is proportional to the nonrelativistic quadrupole moment of the collapsed configuration. We study this spacetime and determine $q_{+}$ such that for $q$, $0 < q < q_{+}$, the outer spacetime singularity remains a null hypersurface. In this case, $\delta$-Kerr spacetime represents a generalized black hole, namely, an asymptotically flat, stationary and axisymmetric vacuum solution of general relativity for which the outer singularity is a closed null hypersurface. For an approximate variant of $\delta$-Kerr spacetime characterized by mass $M$, quadrupole parameter $q$ and angular momentum parameter $a$, where the latter two parameters are treated to first and second orders of approximation, respectively, we analytically determine the quasinormal mode (QNM)  frequencies in the ray approach using the light-ring method as well as in the complementary wave approach for massless scalar field perturbations in the $a = 0$ limit. The QNM frequencies of $\delta$-Kerr spacetime turn out to be nearly the same as  those of the rotating Hartle-Thorne spacetime. 
\end{abstract}

\maketitle

\section{Introduction}

What is the end result of the complete gravitational collapse of neutral matter? Extremely compact gravitationally collapsed systems definitely exist, but the observational evidence regarding their precise nature is thus far inconclusive. We therefore turn to the theoretical evidence based on Einstein's general relativity (GR). The exterior vacuum gravitational field of an electrically neutral stationary asymptotically flat astronomical system is uniquely characterized by a set of exterior multipole moments that do not all generally vanish as $c \to \infty$. Henceforth, for convenience we refer to such moments as nonrelativistic multipole moments. Let us imagine that such a system undergoes complete gravitational collapse.  The spacetime singularity resulting from the complete gravitational collapse is either surrounded by a regular closed null hypersurface that acts as the event horizon, or it is not. In the former case, we have Kerr black holes~\cite{Chandra}, each only characterized by its mass $M$ and angular momentum $J = M\,a\,c$ such that $|a| \le GM/c^2$, while in the latter case the singularity can essentially have arbitrary  gravitoelectric and gravitomagnetic multipole moments. We should mention here in passing that a Kerr black hole does have a \emph{relativistic} quadrupole moment given by $J^2/(Mc^2)$, which vanishes in the Newtonian limit as $c \to \infty$. To form a black hole, the collapsing system must radiate away all of its nonrelativistic quadrupole and higher moments~\cite{Price:1971fb}.

In the case of a Kerr black hole, the requirement of causality is satisfied for physical phenomena occurring in the spacetime region exterior to the event horizon as this null hypersurface acts as a one-way membrane. On the other hand, in \emph{generalized Kerr spacetimes}~\cite{Quevedo:1991zz}, containing various multipole moments, the outer Kerr horizon becomes a naked singularity~\cite{Joshi:2015uoq}.  This is in fact the first singularity that is encountered by radially ingoing observers that start from the asymptotic domain.  

Imagine now a spectrum of asymptotically flat, stationary and axisymmetric Ricci-flat solutions of GR representing collapsed configurations. At one end of the spectrum are Kerr black holes and at the other end generalized Kerr spacetimes. As one moves away from black holes and adds \emph{infinitesimally small} amounts of nonrelativistic higher moments, the horizon becomes singular by the black hole uniqueness theorems; however, we expect via continuity that \emph{the outer singularity would still remain a null hypersurface}. For large values of the multipole moments, the outer singularity will no longer remain a null hypersurface except perhaps for a set of measure zero. Therefore, for sufficiently small magnitudes of higher multipole moments the singular hypersurface still acts as a one-way membrane and causality requirements can be satisfied in the exterior region. The collapsed configurations with these properties that occupy the intermediate region of the spectrum next to black holes are defined to be \emph{generalized black holes}. Generalized black holes were first investigated in our recent paper~\cite{Allahyari:2018cmg} and will be further studied in the present work. 

Most of the recent gravitational wave observations have been interpreted in terms of binary black hole mergers~\cite{Glampedakis:2018blj,Cardoso:2019rvt}. The merging process involves, among other things, the emission of gravitational radiation dominated at late times by damped oscillations characteristic of the final collapsed configuration. This gravitational ringdown involves a spectrum of quasinormal modes (QNMs). Can one extend the theory of black hole QNMs to generalized black holes? Linear  perturbations of the exterior spacetimes of generalized black holes are expected to exhibit damped oscillations characteristic of QNMs, since the appropriate QNM boundary conditions can be imposed due to the existence of the (singular) null hypersurface. We expect that  the spectrum of QNM oscillations of generalized black holes would depend on the higher multipole moments as well. Thus QNMs can be employed to differentiate observationally between generalized black holes and regular black holes.  

In our recent paper~\cite{Allahyari:2018cmg}, we initiated the analytic calculation of the QNMs of generalized black holes with sufficiently small quadrupolar deformations. In particular, we have employed the light-ring method to calculate in the eikonal limit the QNMs of the rotating Hartle-Thorne spacetime~\cite{HaTh} that contains mass $M$, quadrupole moment $Q$ and angular momentum $J$, where the latter two parameters are treated to first and second orders of approximation, respectively.  In our previous work, we showed that  one cannot observationally distinguish a generalized Hartle-Thorne black hole with nonrelativistic quadrupole moment from a Kerr black hole by using ringdown frequencies, provided the dimensionless parameter given by $|QMc^2-J^2|c^2/(G^2M^4)$ is sufficiently small compared to unity. In this paper, we extend  our previous study of quasinormal mode frequencies for generalized black hole solutions to the $\delta$-Kerr spacetime. 

The extension of black holes by the inclusion of higher multipole moments amounts to \emph{deformation} of black holes. We are interested in the simplest case, namely, quadrupolar deformation and have studied the $\delta$-metric~\cite{Allahyari:2018cmg}, which is a deformed Schwarzschild metric given by
\begin{align}\label{I1}
ds^2 = {}&- \left(1-\frac{2\, m}{r}\right)^\delta \,dt^2 \nonumber     \\
{}&+ \left(1-\frac{2\, m}{r}\right)^{1-\delta}\left[\left(\frac{r^2 - 2\,mr}{r^2 - 2\,mr + m^2\,\sin^2\theta}\right)^{\delta^2-1}\left(\frac{dr^2}{1 - \frac{2\,m}{r}}+ r^2 d\theta^2\right)+ r^2\sin^2\theta\,d\phi^2\right]\,.
\end{align}
Henceforth, we assume that $\delta > 0$, unless indicated otherwise. For $\delta = 1$, the static $\delta$-metric reduces to the Schwarzschild metric; therefore, we define $\delta = 1 + q$, where $q$ is proportional to the quadrupole moment of the deformed Schwarzschild metric. We have an oblate configuration for $q > 0$ and a prolate configuration for $-1 < q < 0$.  
With $\delta$ replaced by  $1 + q$ in Eq.~\eqref{I1}, we get the $q$-metric. The $\delta$-metric is generally of Petrov type I; however, it becomes simpler and of type D on the axis of symmetry and can become algebraically special on certain hypersurfaces, see Appendix B of Ref.~\cite{ Allahyari:2018cmg}.  

Metric~\eqref{I1} is known by various names in GR and has been the subject of numerous investigations, see Refs.~\cite{Weyl, Zip, Voo, Papadopoulos:1981wr, Boshkayev:2015jaa,Toshmatov:2019qih,Toshmatov:2019bda} and the references cited therein. The study of the scalar wave perturbations of this $\delta$-spacetime is initiated here in Section VII. The singularities of the $\delta$-metric occur for $r \le 2\,m$; the outer singularity at $r = 2\,m$ is either  a  timelike or a null hypersurface, the latter occurs for oblate deformations up to $q \approx 0.6$~\cite{ Allahyari:2018cmg}. The $r = 2\,m$ singularity is the static limit, where the timelike Killing vector $\partial_t$ becomes null. It is also an infinite redshift surface for outgoing null geodesics~\cite{ Allahyari:2018cmg}. 

The directional nature of the outer naked singularity of  $\delta$-spacetime is modified by the presence of rotation. We study $\delta$-Kerr spacetime and its outer singularity in Sections II--V and then turn to the investigation of the QNM perturbations of these spacetimes. 
In our previous work~\cite{Allahyari:2018cmg}, we studied, among other things, the QNMs of an approximate form of the static $\delta$-metric using the ray picture. In Section VI, we extend our ray analysis to the QNMs of the $\delta$-Kerr solution and compare the results with those of the Hartle-Thorne solution~\cite{ Allahyari:2018cmg}.  Furthermore,
the complementary wave picture is developed in Section VII for massless scalar field perturbations of the exterior $\delta$-metric. This is a novel method of calculating the same QNM frequencies independently via the wave method.  Throughout this paper, Greek indices run from $0$ to $3$, while Latin indices run from $1$ to $3$. The signature of the spacetime metric is $+2$ and units are chosen such that $c = G = 1$, unless specified otherwise.

\section{$\delta$-Kerr Metric}

The $\delta$-Kerr spacetime is a generalized Kerr spacetime; that is, it belongs to the class of Ricci-flat solutions of GR discussed in Ref.~\cite{Quevedo:1991zz}. In recent papers~\cite{Toktarbay:2015lua, Frutos-Alfaro:2016arb}, the rotating $\delta$-metric and its relativistic multipole moments have been considered. To find the $\delta$-Kerr metric, one ``rotates"  the $\delta$-metric via solution generating techniques involving the HKX transformations~\cite{SKMH, Griffiths:2009dfa}; then,  with an appropriate choice of parameters, the result can be expressed as the $\delta$-Kerr metric. To present this metric here, let us first observe that the general stationary axisymmetric  line element in prolate spheroidal coordinates $(t, x, y, \phi)$ can be written  as
\begin{align}\label{M1}
\nonumber ds^2 = {}& - F \,(d t - \omega\, d \phi)^2 \\
{}& + \frac{\sigma^2}{F}\,\left[e^{2\, \gamma} (x^2 - y^2) \left(\frac{d x^2}{x^2 - 1}
+ \frac{d y^2}{1 - y^2} \right) + (x^2 - 1)(1 - y^2)\, d \phi^2 \right]\,,
\end{align}
where  $\sigma > 0$ is a constant length and $F$, $\omega$ and $\gamma$ are functions of $x$ and $y$. Using the static $\delta$-metric with mass parameter $m$ as the seed metric, the explicit expressions for the $\delta$-Kerr functions are given by~\cite{Toktarbay:2015lua,Frutos-Alfaro:2016arb}
\begin{equation}\label{M2}
F = \frac{\mathcal{A}}{\mathcal{B}}\,, \qquad \omega  =   2 \left(a - \sigma\, \frac{\mathcal{C}}{\mathcal{A}} \right)\,
\end{equation}
and
\begin{equation}\label{M3}
e^{2 \gamma} = \frac{1}{4} \left(1 + \frac{m}{\sigma} \right)^{2}\,\frac{\mathcal{A}}{(x^{2} - 1)^{\delta}} \,{\left(\frac{x^{2} - 1}{x^{2} - y^{2}} \right)}^{\delta^2}\,.
\end{equation}
Here, $a$ is the rotation parameter and we assume $|a| \le m$, so that
\begin{equation}\label{M4}
\sigma = \sqrt{m^2-a^2}\,.
\end{equation}
It proves convenient to define a new quadrupole parameter $q$ and a new rotation parameter $\alpha$, $0 \le \alpha \le1$, given by
\begin{equation}\label{M5}
q := \delta - 1\,, \qquad \alpha:= \frac{m-\sigma}{a} = \frac{a}{m+\sigma}\,.
\end{equation}
In terms of these parameters, we define functions $\lambda(x, y)$ and $\eta(x, y)$  such that
\begin{equation}\label{M6}
\lambda  =  \alpha\, (x^{2} - 1)^{- q}\, (x + y)^{2 q}\,, \qquad \eta  =  \alpha \,(x^{2} - 1)^{- q}\, (x - y)^{2 q}\,.
\end{equation}
Furthermore, we define 
\begin{eqnarray}\label{M7}
a_{\pm} & = & (x \pm 1)^{q} 
[x (1 - \lambda \eta) \pm (1 + \lambda \eta)] , \nonumber \\
b_{\pm} & = & (x \pm 1)^{q} [y (\lambda + \eta) \mp (\lambda - \eta)]\,; 
\end{eqnarray}
then, the quantities  $\mathcal{A}$, $\mathcal{B}$ and $\mathcal{C}$ that appear in the metric functions can be expressed as
\begin{eqnarray}\label{M8}
\mathcal{A} & = & a_{+} a_{-} + b_{+} b_{-}\, , \nonumber \\ 
\mathcal{B} & = & a_{+}^{2} + b_{+}^{2}\, , \nonumber  \\
\mathcal{C} & = & \! (x + 1)^{q} \left[x (1 - y^{2})(\lambda + \eta) a_{+} + y (x^{2} - 1)(1 - \lambda \eta) b_{+} \right]\,.   
\end{eqnarray}
Finally, we consider the transformation from the oblate spheroidal coordinates to the Boyer-Lindquist type coordinates, $(t, x, y, \phi) \mapsto (t, r, \theta, \phi)$, with
\begin{equation}\label{M9}
x = \frac{r-m}{\sigma}\,, \qquad y = \cos\theta\,,
\end{equation}
such that metric~\eqref{M1} for the $\delta$-Kerr spacetime takes the form
\begin{align}\label{M10}
\nonumber ds^2 ={}&-F\,dt^2+2\,F\,\omega \,dt\,d\phi+\frac{e^{2 \,\gamma}}{F}\,\frac{\mathbb{B}}{\mathbb{A}}\, dr^2+r^2\,\frac{e^{2 \,\gamma}}{F}\,\mathbb{B}\,d\theta^2\\
{}& + \left(\frac{r^2}{F}\,\mathbb{A}\,\sin^2\theta-F\,\omega^2 \right)d\phi^2\,,
\end{align}
where
\begin{equation}\label{M11}
\mathbb{A} =1-\frac{2\, m}{r}+\frac{a^2}{r^2}\,, \qquad \mathbb{B} = \mathbb{A} +\frac{\sigma^2\,\sin^2\theta}{r^2}\,.
\end{equation}

This Ricci-flat solution of GR has two commuting Killing vector fields $\partial_t$, $\partial_\phi$ and is asymptotically flat. Indeed, for $r \to \infty$, $\lambda \to \alpha$, $\eta \to \alpha$ and one can show that $F \to 1$, $\omega \to 0$ and $\gamma \to 0$, so that metric~\eqref{M10} tends asymptotically to Minkowski metric in spherical polar coordinates. In this connection, it is useful to note that
\begin{equation}\label{M11a}
1 + \alpha^2 = 2\,\frac{m}{a} \,\alpha\,, \qquad 1 - \alpha^2 = 2\,\frac{\sigma}{a} \,\alpha\,.
\end{equation}
The $\delta$-Kerr solution is mirror symmetric about the equatorial plane; that is, for $y \to -y$, $\lambda$ and $\eta$ are switched and $b_{\pm} \to -b_{\pm}$, while the metric functions remain invariant. Furthermore, the axis of rotation  of the $\delta$-Kerr solution is regular (i.e., elementary flat), which follows from a detailed analysis involving the circumstance that along the rotation axis, $\cos \theta = \pm \,1$, $\lambda\,\eta = \alpha^2$, $\gamma = 0$ and $\omega =0$. 

Equation~\eqref{M10} represents the nonlinear superposition of the $\delta$-metric and the Kerr metric. Let us first check that in the absence of rotation we recover the $\delta$-metric. Indeed, for $a = 0$, $\alpha = \lambda = \eta = 0$, $\sigma = m$, $a_{\pm} = (x\pm1)^\delta$ and $b_{\pm} = 0$. Then, 
\begin{equation}\label{M12}
\mathcal{A} = \mathcal{B}\,\left(1-\frac{2\, m}{r}\right)^\delta\,, \qquad \mathcal{B} = \left(\frac{r}{m}\right)^{2\,\delta}\,,  \qquad \mathcal{C} = 0\,
\end{equation}
and
\begin{equation}\label{M13}
F = \left(1-\frac{2\, m}{r}\right)^\delta\,, \qquad \omega = 0\,, \qquad e^{2\,\gamma} = \left(\frac{r^2-2\,mr}{r^2-2\,mr +m^2\,\sin^2\theta}\right)^{\delta^2}\,,
\end{equation}
so that metric~\eqref{M10} reduces to the $\delta$-metric~\eqref{I1}. Let us next suppose that $q = 0$, so that the Schwarzschild metric is now subject to rotation. We find that $\lambda = \eta = \alpha = (m-\sigma)/a$, $a_{\pm} = 2\,\alpha (r-m\pm m)/a$ and $b_{\pm} = 2\,\alpha \, \cos \theta$. Hence, with $\Sigma := r^2 + a^2 \cos^2\theta$, we find
\begin{equation}\label{M14}
\mathcal{A} = \frac{4\,\alpha^2}{a^2}\, (\Sigma - 2m r)\,, \qquad \mathcal{B} = \frac{4\,\alpha^2}{a^2}\,\Sigma\,, \qquad  \mathcal{C} = \frac{4\,\alpha^2}{a\,\sigma}\, [\Sigma - m r(1+\cos^2 \theta)]\,
\end{equation}
and
\begin{equation}\label{M15}
F = 1 - \frac{2\,mr}{\Sigma}\,, \qquad \omega = -2\,ma \,\frac{r\,\sin^2\theta}{\Sigma - 2\,mr}\,, \qquad e^{2\,\gamma} = \frac{\Sigma - 2\,mr}{\Sigma - 2\,mr + m^2\,\sin^2\theta}\,.
\end{equation}
In this case, metric~\eqref{M10} reduces to the standard form of the Kerr metric in Boyer-Lindquist coordinates~\cite{Chandra}, namely,
\begin{equation}\label{M16}
ds^2 = -dt^2+\frac{\Sigma}{\Delta}dr^2+\Sigma\, d\theta^2 +(r^2+a^2)\sin^2\theta\, d\phi^2+\frac{2mr}{\Sigma}(dt-a\sin^2\theta\, d\phi)^2\,,
\end{equation}
where $\Delta=r^2-2mr+a^2$ and $J = m\, a > 0$ is the proper angular momentum of the Kerr source about the $z$ axis.

Finally, we must investigate the limiting case where $a = m$ and $\sigma = 0$. For generalized Kerr spacetimes, one finds in this limit the \emph{extreme} Kerr spacetime regardless of the Newtonian multipole moments~\cite{Quevedo:1991zz}. This is indeed the case for $\delta$-Kerr spacetime as well regardless of the value of $\delta$. To see this, let us first note that as $\sigma \to 0$, Eq.~\eqref{M6} implies
\begin{equation}\label{M17}
\lambda = 1 + \left(2\,\frac{q\,\cos\theta}{r-m} - \frac{1}{m}\right) \,\sigma + O(\sigma^2)\,, \quad  \eta = 1 - \left(2\,\frac{q\,\cos\theta}{r-m} + \frac{1}{m}\right) \,\sigma + O(\sigma^2)\,.
\end{equation}
Next, after some algebra we find $\mathcal{A} = (r^2-2mr+m^2\cos^2\theta)\,\mathcal{D}$, $\mathcal{B} = (r^2+m^2\cos^2\theta)\,\mathcal{D}$ and $\sigma\,\mathcal{C} = m (r-m)(r-m\cos^2\theta)\,\mathcal{D}$,  where
\begin{equation}\label{M18}
\mathcal{D} = \frac{4}{m^2}\,\frac{(r-m)^{2q}}{\sigma^{2q}} \left[1+O(\sigma)\right]\,.
\end{equation}
The metric functions are then obtained from Eqs.~\eqref{M2} and~\eqref{M3}; in this way,  we recover the extreme Kerr spacetime as $a \to m$.

\section{Exterior $\delta$-Kerr Spacetime}

We are interested in the astrophysical aspects of the exterior $\delta$-Kerr spacetime and its perturbations. This spacetime is an asymptotically flat generalized Kerr spacetime~\cite{Quevedo:1991zz}. It is known that for generalized Kerr spacetimes, as one approaches the source from the exterior region, one encounters a singular surface given by the exterior Kerr horizon $m + \sigma$. The singularities are thus confined to the region $r \le m+\sigma$~\cite{Quevedo:1991zz}. To investigate this situation for $\delta$-Kerr spacetime further, we study  the Kretschmann scalar $K$, 
\begin{equation}\label{S1}
K = R_{\mu\nu\alpha\beta}R^{\mu\nu\alpha\beta}\,,
\end{equation}
for the $\delta$-Kerr metric. The function $K$ can be calculated analytically in this case using an algebraic manipulation system; however, the resulting expression is too long and complicated to be manageable. We must therefore resort to numerical analysis; to simplify matters, we concentrate on the \emph{outer} singularity.  A more complete study of the singularity structure of $\delta$-Kerr spacetime is beyond the scope of this paper.

\subsection{Numerical Determination of Outer Singularity}

The analytic  expression for $K$ contains parameters $(m, a, q)$ as well as coordinates $(r, \theta)$. For a given set of parameter values and a fixed value for $\theta$, we can investigate numerically the analytic expression for $K$ as a function of the radial coordinate $r$. In particular,  $\delta$-Kerr solution is symmetric about the equatorial plane ($\theta = \pi/2$) and is asymptotically flat; therefore, $K(r, \theta)$ is symmetric about $\theta = \pi/2$ and $K(r, \theta) \to 0$ as $r \to \infty$. Moving inward along a fixed radial direction from infinity toward the source, we can determine numerically where $K$ first diverges. In this way, the outer singularity of the $\delta$-Kerr spacetime can be determined. The result of our numerical work confirms that the outer singularity occurs at 
\begin{equation}\label{S2}
r_S = m + \sigma\,,
\end{equation}
which is the outer horizon of the Kerr metric. This result is in agreement with the theorems regarding the uniqueness of the black hole solutions of general relativity. 

It is interesting to note that for $q = 0$, $K(r, \theta)$ reduces to its Kerr value, namely, 
\begin{equation}\label{S3}
K_{\rm Kerr}= 48\,m^2\,\frac{r^2-a^2\cos^2\theta}{(r^2+a^2\cos^2\theta)^{6}}\,(r^2 - 4\,r a \cos\theta+a^2\cos^2\theta)(r^2 +4\,r a \cos\theta+a^2\cos^2\theta)\,.
\end{equation}
In this expression, $K$ diverges at $r = 0$ along the $\theta=\pi/2$ direction (i.e.,  in the equatorial plane), as expected for the ring singularity of Kerr spacetime. Moreover,  for the Kerr-Q spacetime, which is an approximate variant of the $\delta$-Kerr spacetime valid to first order in $q$ and second order in $a$, the analytic expression for $K$ is given  in Eq.~\eqref{W11a} of  Section V and its implications are explored.

\begin{figure}
\includegraphics[scale=0.4]{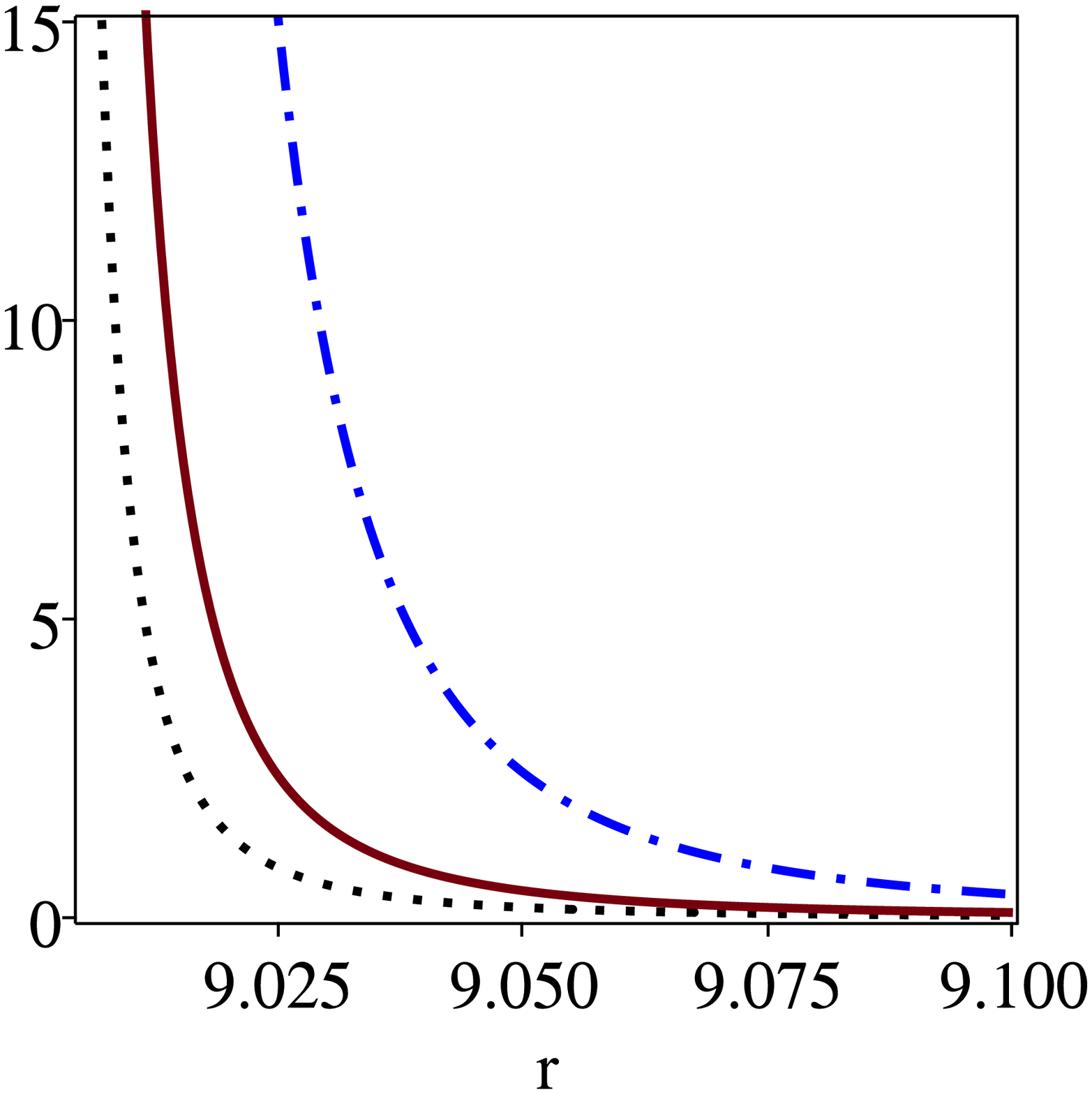}
\hspace{0.5cm}
\includegraphics[scale=0.4]{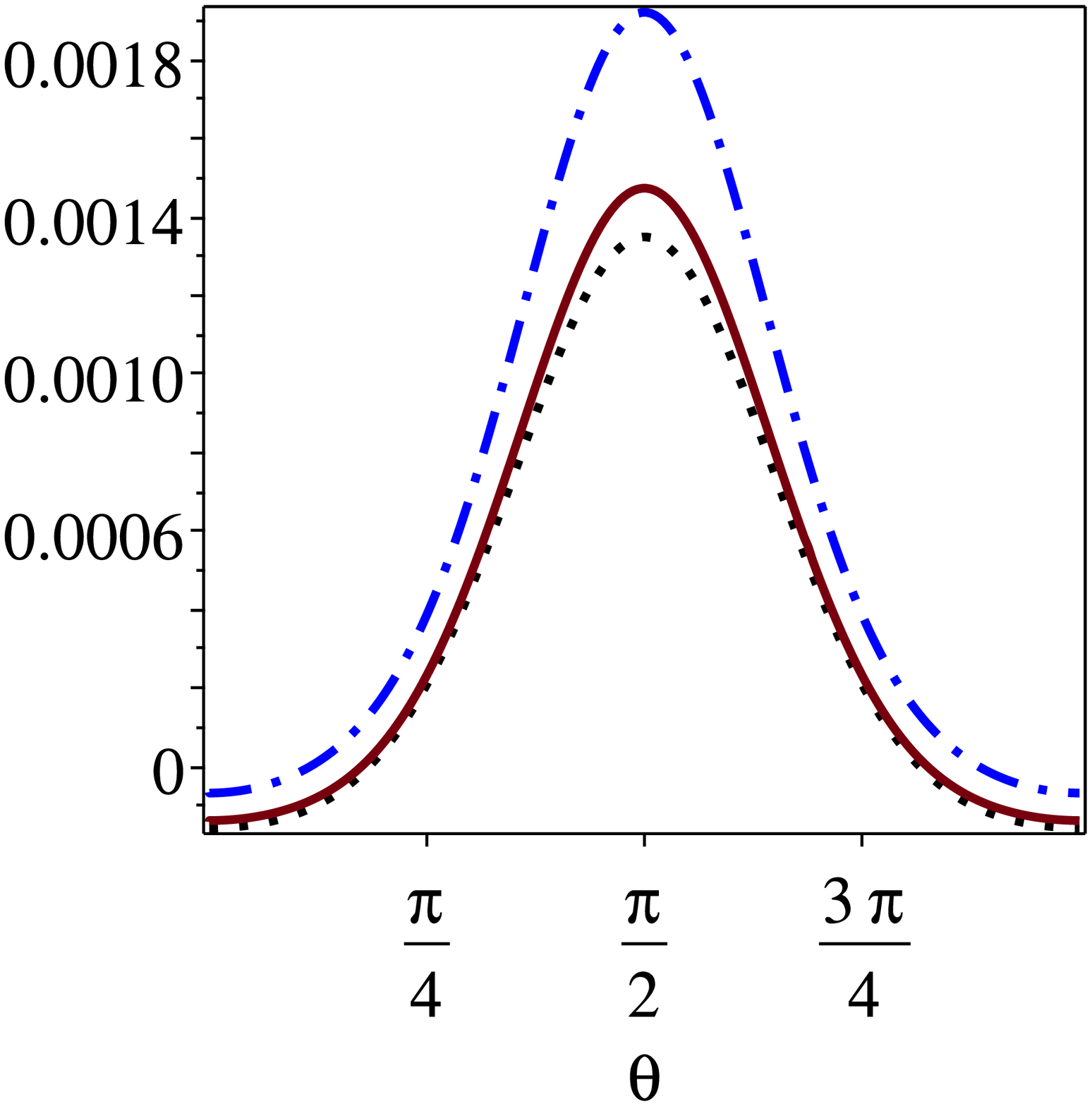}
\caption{\label{fig1} 
The behavior of the Kretschmann invariant $K(r, \theta)$ for $m=5$, $a=3$ and $q=1/8$ (black dot), $q=1/6$ (solid red) and $q=1/4$ (blue dot-dash).  Left panel: $K$ as a function of $r: 9 \to \infty$ for $\theta=\pi/2$. The singularity of $K$ occurs at the outer Kerr horizon $r_S=9$. Right panel: $K$ as a function of $\theta : 0 \to \pi$ for $r=10$.}
\end{figure}

\begin{figure}
\includegraphics[scale=0.4]{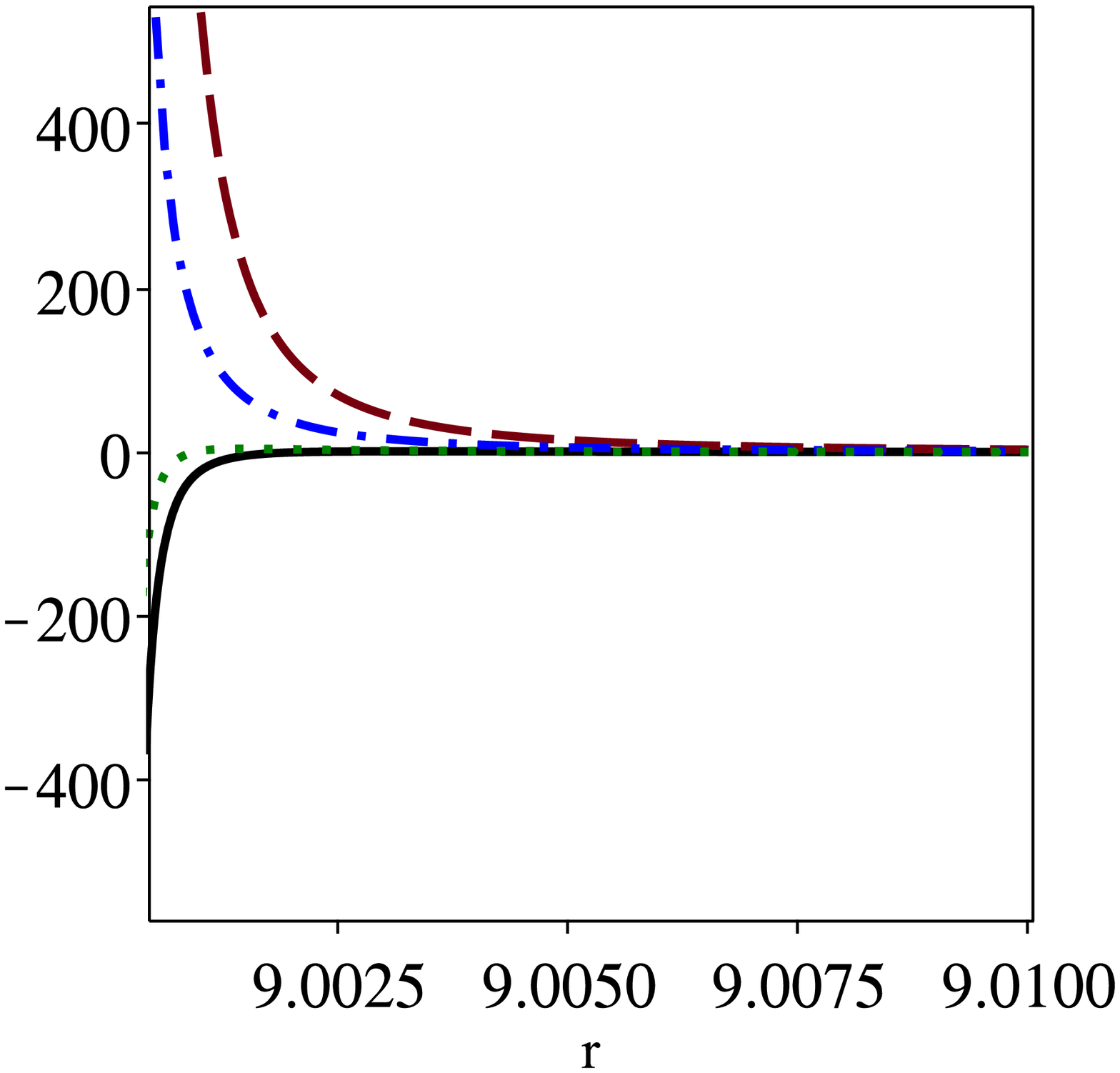}
\hspace{0.5cm}
\includegraphics[scale=0.4]{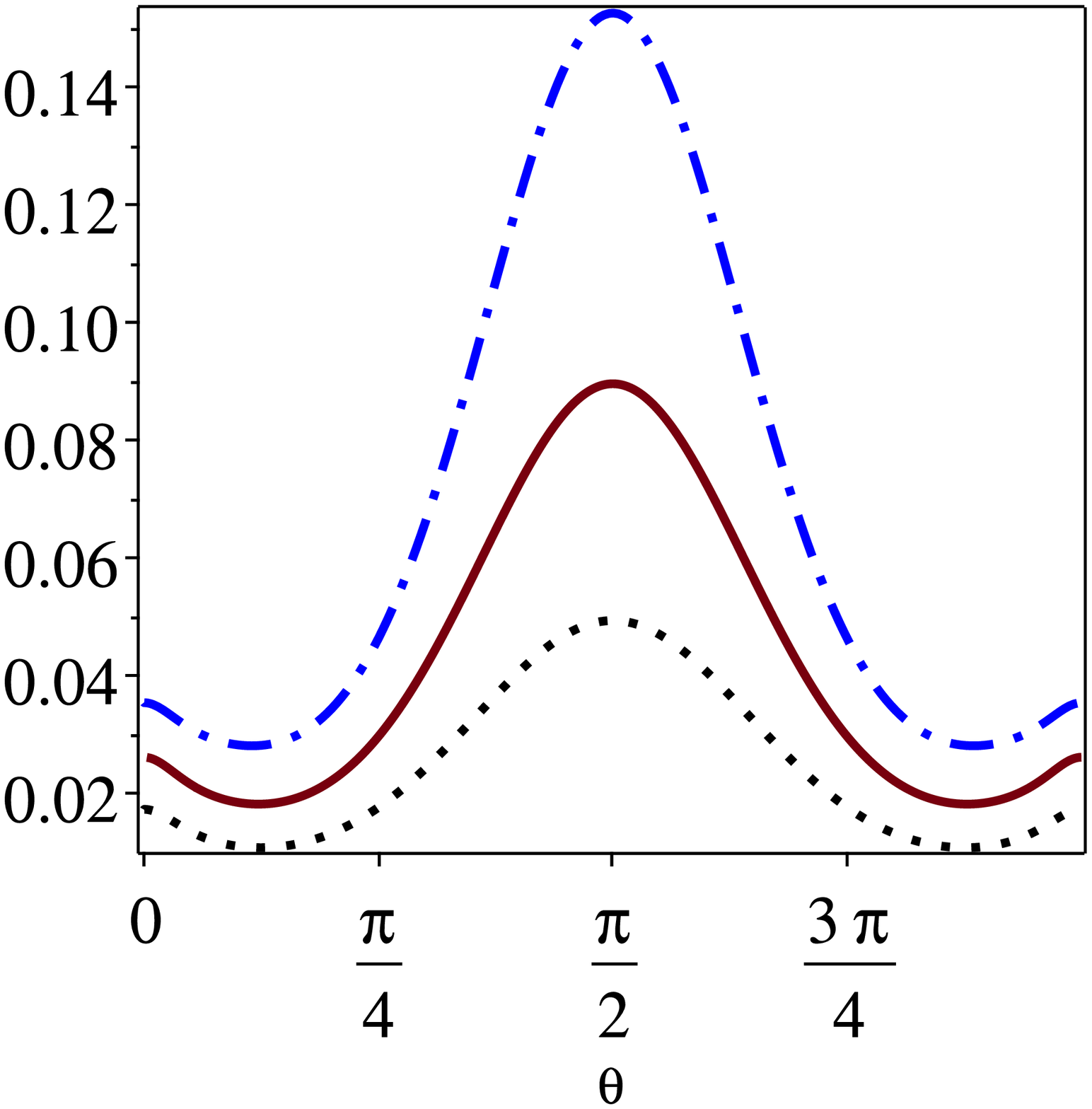}
\caption{\label{fig2} 
The behavior of the Kretschmann invariant $K(r, \theta)$ for $m = 5$ and $a = 3$ near the outer singularity $r_S=9$.  Left panel: $K$ as a function of $r: 9 \to \infty$ for $q = 0.1$ and  $\theta=\pi/10$ (green dot), $\theta = \pi/6$ (solid black), $\theta = \pi/3$ (blue dot-dash) and $\theta = \pi/2$ (red dash). We have checked numerically that these results remain essentially the same for various values of $q$. Right panel: $K$ as a function of $\theta : 0 \to \pi$ for $r = 9.05$ and $q = 0.08$ (black dot), $q = 0.1$ (solid red) and $q = 0.12$ (blue dot-dash).}
\end{figure}

For oblate configurations under consideration in this paper, $q > 0$ and numerical work reveals that as expected $K$ vanishes asymptotically as $r \to \infty$. Moreover,  depending on the values of $r$ and $\theta$, $K$ can be positive or negative.  For our numerical studies of $K(r, \theta)$, we first assume that in units of a certain fixed measure of length, $m=5$ and $a=3$, so that $\sigma=4$ and $\alpha= 1/3$. In Figures~\ref{fig1} and~\ref{fig2}, we present the behavior of $K$ for different values of $q$. 
 The left panel of Fig.~\ref{fig1} contains the plot of $K$ versus $r$ in the equatorial plane ($\theta=\pi/2$) for $q = 1/8, 1/6$ and $1/4$, and we note that $K$ is positive and diverges at the outer Kerr horizon, Eq.~\eqref{S2}, given in this case by $r_S = 9$ and rapidly drops to zero as $r \to \infty$. The right panel of Fig.~\ref{fig1} presents the plot of $K$ as function of $\theta$ while $r$ is held fixed at $r=10$.  In this case, $K(r = 10, \theta)$ is symmetric about the equatorial plane and $K > 0$ near the equatorial plane, but turns negative near the poles. 

Let us now consider $K(r, \theta)$ for the same fixed $m$ and $a < m$ as in Figure 1. Let us assume that  $r$ is very close to the outer singularity $r_S$. Our numerical results, presented in the left panel of Figure 2,  indicate that as $r$ approaches $r_S$, $K$ diverges and either approaches $ +\infty$ or $-\infty$ depending on the polar angle $\theta$.  However, as we radially move away from the outer singularity, $K$ can be positive for all values of $\theta$ as indicated in the right panel of Figure 2 for $r = 9.05$ and $q = 0.08$, $0.1$ and $0.12$.  In fact, for fixed $\theta$ near the poles, the behavior of $K$ could be oscillatory; in any case, it is evident from the comparison of the two figures that as one gets closer to the $r_S = 9$ singularity, the behavior of $K(r, \theta)$ as a function of $\theta$ becomes more complex.

\subsection{Nature of the Outer Singularity}

The outer event horizon of Kerr spacetime with radial coordinate $m + \sigma$ is a regular null hypersurface. In $\delta$-Kerr spacetime, however, it becomes singular and is in fact the outer singularity of $\delta$-Kerr spacetime. Is it still null? To investigate this question, let $N$ be the vector normal to the  $r=$ constant hypersurface; then, its norm is given by 
\begin{align}\label{S4}
N\cdot N = g^{rr}=  e^{-2 \gamma}\,F\,\frac{\mathbb{A}}{\mathbb{B}}.
\end{align}
To compute this expression, it is convenient to note that
\begin{equation}\label{S5}
x^2 -1= \frac{r^2}{\sigma^2} \,\mathbb{A}\,, \qquad x^2 -y^2 = \frac{r^2}{\sigma^2} \,\mathbb{B}\,.
\end{equation}
Moreover, it is useful to introduce
\begin{equation} \label{S6}
\mathbb{C} =\frac{r-m}{\sigma}+\cos\theta\,, \qquad \mathbb{D} =\frac{r-m}{\sigma}-\cos\theta\,, \qquad  \mathbb{C}\,\mathbb{D} =\frac{r^2}{\sigma^2}\,\mathbb{B}\,.
\end{equation}
It is then staightforward to write
\begin{equation}\label{S7}
N \cdot N=\frac{A}{B}\,,
\end{equation}
where
\begin{equation}\label{S8}
A = \frac{4\,\sigma^2}{(m+\sigma)^2}\,\left(1-\frac{m -\sigma}{r}\right)^{2-2q}\, \mathbb{B}^{2q+q^2}\,\mathbb{A}^{-1+3q-q^2}\,
\end{equation}
and
\begin{align}\label{S9}
B = {}&\alpha^2\,\left(\frac{\sigma}{r}\right)^{4q+2}\,\left(1-\frac{m -\sigma}{r}\right)^2\,\mathbb{A}^{2q-2}\left[\left(1+\cos\theta \right)\mathbb{D}^{2q}-\left(1-\cos\theta \right)\mathbb{C}^{2q} \right] ^2 \nonumber\\
{} &+\left[\left(1-\frac{m -\sigma}{r}\right)^2\,\mathbb{A}^{2q -1} -\alpha^2 \,\mathbb{B}^{2q}  \right]^2\,. 
\end{align}
For a null hypersurface, we must have $N\cdot N=0$. We recall that $\mathbb{A} = 0$ for $r_{\pm}=m\pm\sigma$, which determine the horizons for Kerr spacetime. A detailed analysis reveals that in  
$\delta$-Kerr spacetime with $\theta \ne 0,\pi$,  the singular hypersurface $r_S =m+\sigma$  is null if we have  $-1+3q-q^2 > 0$. Thus $q_{-} < q < q_{+}$, where $q_{\pm} = (3 \pm \sqrt{5})/2$.

We are interested in oblate configurations with $q>0$. Therefore, the outer singularity at $r_S=m+\sigma$ is a null hypersurface for $0 < q < (3 + \sqrt{5})/2$.  Along the rotation axis, where $\theta = 0,\pi$, we have $\mathbb{B} =\mathbb{A}$; in this case,  we find that the condition $1+q > 0$ is always satisfied by definition, so that the outer singularity is always null. 

It is interesting to compare our present result with that for the static $\delta$-metric, where the singular hypersurface at $r_S = 2\,m$ is null for $0 < q < (\sqrt{5} -1)/2$~\cite{Allahyari:2018cmg}. There appears to be no explicit dependence of the maximum allowed value of $q$ upon the rotation parameter $a$. To explain how this comes about, let us first observe that the outer singularity occurs where $\mathbb{A} = 0$; moreover, due to the specific form of the $\delta$-Kerr metric, the exponents of $\mathbb{A}$ in the metric coefficients depend \emph{only} upon $q$. However, the functional form of Eq.~\eqref{S7} depends, among other things, upon the rotation parameter $a$. Indeed, Eq.~\eqref{S7} has different functional forms involving $\mathbb{A}$ for $a=0$ and $a\ne 0$. In effect, the outer singularity is null for $1-q-q^2 > 0$ and $-1+3q-q^2 > 0$ if $a = 0$ and $a \ne 0$, respectively. To see this in more detail, we note that 
in the absence of rotation ($a = 0$), $\alpha = 0$ and this changes the functional form of Eq.~\eqref{S7} since Eq.~\eqref{S9} reduces to $B = \mathbb{A}^{4q-2}$. We find in this case, $N\cdot N = \mathbb{B}^{2q+q^2}\,\mathbb{A}^{1-q-q^2}$, so that the outer singularity is null for $1-q-q^2 > 0$, which is our previous result~\cite{Allahyari:2018cmg}. Therefore, the mere fact that the source rotates ($a \ne 0$)  increases the corresponding range of $q$ for which the outer singularity is null and this circumstance is independent of the magnitude of the nonzero rotation parameter $a$, $ 0 < a \le m$.   For the rotating source, the maximum allowed value of $q$ is about four times larger than for the static source, so that the null hypersurface can act as a one-way membrane for considerably larger values of the quadrupole parameter in the $\delta$-Kerr spacetime.

\subsection{Stationary Limit}

The norm of the timelike Killing vector  $\partial_t$ vanishes at the stationary limit surface. Let us recall that the metric function $F(r, \theta) = \mathcal{A} / \mathcal{B}$ is related to the square of the norm of $\partial_t$ such that $\partial_t \cdot\partial_t = g_{tt}=-F$. Thus the stationary limit surface is given by $F(r, \theta) = 0$. The analytic expression for this surface is too complicated to be useful; instead, we can numerically evaluate $F(r, \theta)$  for a fixed $\theta$ along the radial direction from $F(\infty, \theta) = 1$ to zero. In this way, we have verified that for $q=0$, we get Kerr's outer stationary limit surface at $r = m +\sqrt{m^2-a^2\,\cos^2\theta}$; however, for $q > 0$ the outer stationary limit of $\delta$-Kerr spacetime depends explicitly on $q$. This is in contrast to $\delta$-spacetime, where the static limit surface $r = 2\,m$ is independent of $q$~\cite{Allahyari:2018cmg}.

\section{Weak-Field Approximation}

Let us imagine a slowly rotating Schwarzschild gravitational field with a small spheroidal deformation $q > 0$. For sufficiently small deviations from the Schwarzschild metric, we may expand the 
$\delta$-Kerr metric to first order in $q$ and to second order in $a$. We neglect terms proportional to $q \,a^2$ and $q\, a$. The result is,
\begin{equation}\label{W1}
- g_{tt}=\hat{\mathbb{A}}\left(1+q\ln \hat{\mathbb{A}} \right)+\frac{2 a^2 m \cos^2\theta}{r^3}\,, \qquad - g_{t\phi} =\frac{2am\sin^2\theta}{r}\,,
\end{equation}
\begin{equation}\label{W2}
g_{rr}=\frac{1}{\hat{\mathbb{A}}}\left(1+q\ln\frac{\hat{\mathbb{A}}}{\hat{\mathbb{B}}^2} \right) -a^2\frac{1- \hat{\mathbb{A}}\,\cos^2\theta}{r^2\hat{\mathbb{A}}^2}\,, \qquad
g_{\theta\theta}=r^2+a^2\cos^2\theta+qr^2\ln\frac{\hat{\mathbb{A}}}{\hat{\;\mathbb{B}}^2}\,
\end{equation}
and
\begin{equation}\label{W3}
g_{\phi\phi} =\left[ r^2-qr^2\ln\hat{\mathbb{A}}+a^2\left( 1+\frac{2m\sin^2\theta}{r}\right) \right]  \sin^2\theta\,.
\end{equation}
Here,  
\begin{equation}\label{W4}
\hat{\mathbb{A}}:=1-\frac{2m}{r}\,, \qquad \hat{\mathbb{B}}:=1-\frac{2m}{r}+\frac{m^2\sin^2\theta}{r^2}\,.
\end{equation}

This post-Schwarzschild metric represents the \emph{linear} superposition of the approximate forms of the Kerr and $q$ metrics. When $a=0$, we get the $\delta$-metric  to first order in $q$~\cite{Allahyari:2018cmg}, while for $q=0$, we get the Kerr metric to second order in $a$.
To find the appropriate multipole moments for this metric, we need to have its weak-field post-Newtonian limit. Thus, we expand the above metric to third order in $m/r$. The last step consists of finding a coordinate transformation in which the metric takes the post-Newtonian form.
Let us use the transformations
\begin{equation}\label{W5}
r = \rho\,\left[1- q\,\frac{m}{\rho} - q\,\frac{m^2}{\rho^2}\left(1 + \frac{m}{\rho} \right)\,\sin^2\vartheta \right] - \frac{a^2}{2\rho}\, \left(1 +\frac{m}{\rho} + \cdots \right)\,\sin^2\vartheta\,
\end{equation}
and
\begin{equation}\label{W6}
\theta = \vartheta-\left[  q\,\frac{m^2}{\rho^2}\left(1 + 2 \frac{m}{\rho} \right)\, + \frac{a^2}{2\rho^2}\, \left(1 +\frac{2m}{\rho} + \cdots\right)\,\right] \sin\vartheta\,\cos\vartheta\,.
\end{equation}
Expanding in powers of $1/\rho$, we have neglected $a^2m^2/\rho^4$ and higher.  The $\delta$-Kerr metric then takes the form
\begin{align}\label{W7}
ds^2&=-\left( 1+2\Phi\right) dt^2-4\frac{a\,M}{\rho^3}\sin^2\vartheta \,dt\,d\phi+\frac{d\rho^2}{1+2\Phi}+U(\rho,\vartheta)\rho^2\left(d\vartheta^2+\sin^2\vartheta d\phi^2 \right)\,, 
\end{align}
where 
\begin{equation}\label{W8}
\Phi=-\frac{M}{\rho}+\frac{Q}{\rho^3}\,P_2({\cos\vartheta})\,, \qquad U(\rho,\vartheta)=1-\frac{2\,Q}{\rho^3}\,P_2({\cos\vartheta})\,.
\end{equation}
Here, $P_2({\cos\vartheta}) = (3\,\cos^2 \vartheta -1)/2$ is a Legendre polynomial and
\begin{align}\label{W9}
M=m(1+q)\,,\qquad Q=\frac{2}{3}M^3 q + \frac{J^2}{M}\,,\qquad J=M\,a\,.
\end{align}
Comparing our metric with the weak-field post-Newtonian metric in GR (that includes the Lense-Thirring term), we recognize that the solution represents a source with mass $M$, quadrupole moment $Q$ and angular momentum $J$. 

The standard approach to the problem of multipole moments of stationary configurations in general relativity is rather formal and is based on the invariantly defined Geroch-Hansen moments~\cite{Q2,Raposo:2018xkf}. For the application of this formal approach to the rotating $\delta$-metric, see Ref.~\cite{Frutos-Alfaro:2016arb}. We use instead the weak-field approximation, which provides  insight into the correspondence with Newtonian gravity. Our results agree with the invariant definitions in the appropriate limits.  In this way, we can relate the relativistic quadrupole moment to the one usually employed in Newtonian gravity~\cite{Allahyari:2018cmg}.

\section{Kerr-$Q$ Metric}

We are interested in the QNMs of the $\delta$-Kerr spacetime when the quadrupole parameter $q > 0$ is less than the maximum allowed ($\approx 2.6$) in Section III. To simplify matters, we consider the $\delta$-Kerr spacetime to first order in quadrupole moment and second order in angular momentum. In this approximation scheme, it is still possible to access the strong-field regime near the outer singularity. Furthermore, we need the metric to depend on $M$, $q$ and $a$. Taking Eq.~\eqref{W9} into account, we replace $m$ by $M/(1+q)$ in 
 the $\delta$-Kerr metric and expand the resulting metric to first order in $q$ and second order in $a$. In this way, we find the Kerr-$Q$ metric,
\begin{align}\label{W10}
ds_{KQ}^2 ={}& -  \left[\tilde{\mathbb{A}}+q\left(\frac{2M}{r\tilde{\mathbb{A}}} +\ln{\tilde{\mathbb{A}}}\right)\tilde{\mathbb{A}}+\frac{2 a^2 M \cos^2\theta}{r^3}\right]\,dt^2-\frac{4a M\sin^2\theta}{r}dt\,d\phi \nonumber  \\
{}&+ \left[\frac{1}{\tilde{\mathbb{A}}}-\frac{q}{\tilde{\mathbb{A}}}\left(\frac{2M}{r\tilde{\mathbb{A}}} +\ln{\frac{\tilde{\mathbb{B}}^2}{\tilde{\mathbb{A}}}}\right)-a^2\frac{1-\tilde{\mathbb{A}}\,\cos^2\theta}{r^2\tilde{\mathbb{A}}^2}\right]dr^2 \nonumber    \\
{}&+ \left(1-q\ln{\frac{\tilde{\mathbb{B}}^2}{\tilde{\mathbb{A}}}}+\frac{a^2}{r^2}\cos^2\theta\right)\,r^2\,d\theta^2   \nonumber   \\
{}& + \left[  1-q\ln{\tilde{\mathbb{A}}}+\frac{a^2}{r^2}\left( 1+\frac{2M\sin^2\theta}{r}\right) \right]  \,r^2\sin^2\theta\,d\phi^2\,,
\end{align}
where $a = J/M$, $q = 3\,(Q-J^2/M)/(2\,M^3)$ and 
\begin{equation}\label{W11}
\tilde{\mathbb{A}}:=1-\frac{2M}{r}\,, \qquad \tilde{\mathbb{B}}:=1-\frac{2M}{r}+\frac{M^2\sin^2\theta}{r^2}\,.
\end{equation}
This metric reduces to the $SQ$-metric defined in our previous paper~\cite{Allahyari:2018cmg} for $a=0$ and to the Kerr metric (to second order in $a$) when $q=0$. Let us briefly mention here that the $SQ$-metric is the deformed Schwarzschild metric with linear quadrupole moment $Q = 2\,q M^3/3$. While the $\delta$-metric has parameters $m$ and $q$, the $SQ$ metric has parameters $M$ and $q$ and is an approximate variant of the $\delta$-metric where $q$ is taken into account to linear order. 

It is interesting to compute the Kretschmann scalar $K$ for the Kerr-Q metric. The result is
\begin{equation}\label{W11a}
K_{\rm Kerr-Q} = \frac{48\,M^2}{r^6}\left\{1-21\,\frac{a^2}{r^2} \cos^2\theta + 2\,q \left[\ln{\frac{ \tilde{\mathbb{B}}^2}{\tilde{\mathbb{A}}}} - \frac{M}{r\, \tilde{\mathbb{B}}}\left(1-\frac{M}{r}\sin^2\theta\right)\right]\right\}\,,
\end{equation}
where for $q = 0$ we recover Eq.~\eqref{S3} for $K_{\rm Kerr}$ to second order in $a/r$. It is clear from Eq.~\eqref{W11a} that the outer singularity occurs at $r = 2\,M$ such that $K_{\rm Kerr-Q} \to \infty$ for $q > 0$ and $0 < \theta < \pi$. However, along the axis of symmetry,  $K_{\rm Kerr-Q} \to -\infty$ for $q > 0$. The existence of the outer singularity is thus independent of the polar angle $\theta$ in this case.

\section{QNMs of Kerr-$Q$ Spacetime: The Ray Picture}

The response of the collapsed configuration to linear massless (scalar, electromagnetic and gravitational) perturbations of the exterior spacetime domain may be dominated at late times by certain resonant modes in the form of damped oscillations characteristic of the collapsed system. We denote the frequency of such a QNM by $\Omega^0 + i\, \Gamma$. The linear perturbations of exterior generalized black holes satisfy complicated equations that do not appear to be separable, a circumstance that is further discussed in the next section. To avoid the separability problem, we employ the light-ring method  as described in detail in our previous paper~\cite{Allahyari:2018cmg}. This approach, first used in Refs.~\cite{FeMa1, FeMa2, Mashhoon:1985cya} for the calculation of QNMs of Kerr and Kerr-Newman black holes,  is independent of the separability of the massless perturbation equations. On the other hand, only certain QNMs can be determined via this eikonal method. 

The light-ring method involves the analysis of perturbations on the exterior background caused by the initially circular unstable bundles of prograde and retrograde null rays in the equatorial plane of the rotating collapsed configuration. We study the propagation of the rays outward to spatial infinity and inward toward the outer null singularity. The rays of the radiation correspond to the eikonal limit of  waves with high frequency $\Omega \gg M^{-1}$. Moreover, the motion takes place in the equatorial plane, so that the waves have angular momentum parameters $(j, \mu)$ with $|\mu| = j \gg 1$. The prograde (retrograde) waves have $\mu = j$\, ($\mu = - j$). The real parts of the QNM frequencies turn out to be equal to $\pm\, j\, \Omega_{\pm}$, where $\Omega_{+}$ 
($\Omega_{-}$) is the orbital frequency of the circular prograde (retrograde) null rays. The imaginary parts of the QNM frequencies have to do with the decay of the orbits and can be characterized via the corresponding Lyapunov exponents~\cite{Cardoso:2008bp}. For recent work related to the light-ring method, see  Refs.~\cite{Konoplya:2017wot,Assumpcao:2018bka,McWilliams:2018ztb,Glampedakis:2017cgd}, while Refs.~\cite{Ferrari:2007dd,Berti:2009kk,Konoplya:2011qq} contain general review articles regarding QNMs.

The light-ring method that we employ to find the QNM frequencies is such that the result will be a linear superposition of the QNM frequencies of the $SQ$-metric, which we found in our previous paper~\cite{Allahyari:2018cmg} and are further discussed in the next section, and those of the Kerr metric determined via the light-ring method to second order in $a = J/M$~\cite{FeMa2}.  However, to render the present paper essentially self-contained, we briefly present the details of the calculation in Appendices A and B. The resulting QNM frequencies are  
\begin{align}\label{W12}
\Omega^0_{KQ} + i\, \Gamma_{KQ} = {}& \frac{j}{3\sqrt{3}M}\,\left[ 1 - q\, \left( -1+\ln{3}\right)  \pm \frac{2a}{3\sqrt{3}M} +\frac{11}{54}\, \frac{a^2}{M^2}\right]\,,  \nonumber  \\
{}&+ i\,\frac{n + \frac{1}{2}}{3\sqrt{3}M}\,\left\{1 +q\,\left[1 + 2\ln\left(\frac{2}{3}\right)\right]-\frac{2}{27}\,\frac{a^2}{M^2}\right\}\,,\qquad n = 0, 1, 2, 3, \cdots\,,
\end{align}
where $q$ is given by Eq.~\eqref{W9}, namely,
\begin{equation}\label{W13}
q := \frac{3}{2}\,\frac{Q-J^2/M}{M^3}\,.
\end{equation}
When $q = 0$, Eq.~\eqref{W12} reduces to the corresponding Kerr QNM frequencies to second order in $(a/M)^2$, which are described in detail in Appendix E of Ref.~\cite{Allahyari:2018cmg}.

These results should be compared with the QNMs of the rotating Hartle-Thorne solution, which were derived in Ref.~\cite{Allahyari:2018cmg} using the same light-ring method.  That is,
\begin{align}\label{W14}
\Omega^0_{RHT} + i\, \Gamma_{RHT}  = {}& \frac{j}{3\sqrt{3}M}\,\left[1  - \frac{1}{2}\,\tilde{q}\,(-16 +15\ln3) \pm \frac{2J}{3\sqrt{3}M^2} +  \frac{11}{54}\,\frac{J^2}{M^4}\right]  \nonumber   \\
{}&+ i\,\frac{n + \frac{1}{2}}{3\sqrt{3}M}\left[1 +  \tilde{q}\,(-16 + 15\ln{3}) - \frac{2}{27}\,\frac{J^2}{M^4}\right]\,, \qquad n = 0, 1, 2, 3, \cdots\,,
\end{align}
where
\begin{equation}\label{W15}
\tilde{q} := \frac{5}{8}\,\frac{Q-J^2/M}{M^3}\,.
\end{equation}

Comparing the two expressions for the QNM frequencies, we note that the parts involving the quadrupole parameters $q$ and $\tilde{q}$ are indeed different from each other; however, a closer examination reveals that they are almost essentially the same due to the circumstance that
\begin{equation}\label{W16}
\frac{\frac{3}{2}\,\left( -1+\ln{3}\right)}{\frac{5}{16}\,(-16 + 15\ln{3})}\approx 0.988\,, \qquad  \frac{\frac{3}{2}\,\left[1 + 2\ln\left(\frac{2}{3}\right)\right]}{\frac{5}{8}\,(-16 + 15\ln{3})}\approx 0.947\,.
\end{equation}

\section{QNMs of the $\delta$-Spacetime: The Wave Picture}

The main purpose of our investigations in Ref.~\cite{Allahyari:2018cmg} and the present work has been the calculation of QNM frequencies of a black hole with quadrupole moment. The simplest solution of GR involving a deformed Schwarzschild black hole with a quadrupole moment is given by the $\delta$-spacetime. We used the light-ring method that employs the particle (i.e., null ray) picture in Ref.~\cite{Allahyari:2018cmg} to calculate certain QNMs of $\delta$-spacetime to linear order in $q = \delta -1$, namely,
\begin{align}\label{U0}
\Omega_{SQ}  = {}& \frac{j}{3\sqrt{3}M}\,\left[ 1 + q\, \left( 1-\ln{3}\right)\right]\,,  \nonumber  \\
{}&+ i\,\frac{n + \frac{1}{2}}{3\sqrt{3}M}\,\left\{1 +q\,\left[1 + 2\ln\left(\frac{2}{3}\right)\right]\right\}\,,\qquad n = 0, 1, 2, 3, \cdots\,.
\end{align}
This is our main result for $\delta$-spacetime and is the $J = 0$ limit of Eq.~\eqref{W12}; see also Eqs.~(98)--(99) of Ref.~\cite{Allahyari:2018cmg}. 

It is useful to derive Eq.~\eqref{U0} using the \emph{wave picture} by employing the evolution of linear perturbations caused by massless scalar field perturbations on the exterior of $\delta$-spacetime. To this end, we first derive the perturbation equation in this section and then discuss its implications. 

\subsection{Massless Scalar Field on $\delta$-Spacetime Background}

Imagine a linear perturbation of $\delta$-spacetime by a massless scalar field $\Psi$ that satisfies the wave equation
\begin{equation}\label{U1}
\partial_\mu\,\left(\sqrt{-g}\,g^{\mu\nu}\,\partial_\nu  \Psi\right) = 0\,
\end{equation}
and has negligible influence on the background geometry. That is, the energy-momentum tensor of the perturbing field,
\begin{equation}\label{U1a}
T_{\mu \nu} = \partial_\mu \Psi\,\partial_\nu  \Psi - \frac{1}{2}\, g_{\mu \nu} \,g^{\alpha \beta}\, \partial_\alpha \Psi\,\partial_\beta  \Psi\,,
\end{equation}
is a quantity of second order in the scalar field $\Psi$ and hence generates a metric perturbation of $\delta$-spacetime via Einstein's field equation 
\begin{equation}\label{U1b}
R_{\mu \nu} - \frac{1}{2} \,g_{\mu \nu} R = \frac{8 \pi G} {c^4} T_{\mu \nu}\,
\end{equation}
that is of second order and can therefore be ignored in our linear perturbation scheme. 

Let us write $\delta$-metric~\eqref{I1} in the form
\begin{equation}\label{U2}
ds^2 = -\hat{\mathbb{A}}^\delta \,dt^2 + \hat{\mathbb{A}}^{-\delta}\, \left(\frac{\hat{\mathbb{A}}}{\hat{\mathbb{B}}}\right)^{\delta^2-1}\,dr^2 +  \hat{\mathbb{A}}^{1-\delta}\,\left(\frac{\hat{\mathbb{A}}}{\hat{\mathbb{B}}}\right)^{\delta^2-1}\,r^2\,d\theta^2 +\hat{\mathbb{A}}^{1-\delta}\,r^2\,\sin^2\theta\,d\phi^2\,,
\end{equation}
where $\hat{\mathbb{A}} = 1 -2\,m/r$ and $\hat{\mathbb{B}} = \hat{\mathbb{A}} + m^2\,\sin^2\theta/r^2$; see Eq.~\eqref{W4}. Moreover, in this case
\begin{equation}\label{U3}
\sqrt{-g}= \frac{{\hat{\mathbb{A}}^{\delta^2-\delta}}}{{\hat{\mathbb{B}}^{\delta^2-1}}}\, r^2\,\sin\theta\,.
\end{equation}
Let us note that for $\theta \ne 0, \pi$, the volume element corresponding to $r = 2 m$ vanishes for $\delta > 1$, so that $r = 2 m$ acts like the ``origin" for external observers. 

The solution of the scalar wave equation in general depends upon $\delta$. In this connection, we mention that for $\delta = 0$, the exterior spacetime is flat, since metric~\eqref{U2} can be obtained in this case from
\begin{equation}\label{U3a}
ds^2 = - dt^2 + d\rho^2 + \rho^2\, d\phi^2 + dz^2\,,
\end{equation}
via 
\begin{equation}\label{U3b}
\rho = \sqrt{r(r-2m)}\,\sin\theta\,, \qquad z = (r-m)\, \cos\theta\,.
\end{equation}
For $\delta = 1$, metric~\eqref{U2} reduces to the Schwarzschild metric and for $\delta = -1$, metric~\eqref{U2} again reduces to the standard Schwarzschild metric upon formally replacing the radial coordinate $r$ by $2m-r$. 

The $\delta$-spacetime is static and axisymmetric; therefore, we can assume a massless scalar field of the form
\begin{equation}\label{U4}
 \Psi = \frac{1}{r}\, e^{i\,\Omega\, t-i\,\mu\,\phi}\,\psi(r,\theta)\,,
\end{equation}
where the frequency $\Omega$ and multipole parameter $\mu$ are constants. Substituting this ansatz in the wave equation, we find
\begin{align}\label{U5}
{}& r\,\frac{\partial}{\partial r}\left[\left(1-\frac{2\,m}{r}\right) r^2\,\frac{\partial}{\partial r}\left(\frac{\psi}{r}\right)\right] + \frac{1}{\sin\theta}\,\frac{\partial}{\partial \theta}\left(\sin\theta\,\frac{\partial \psi}{\partial \theta}\right) \nonumber    \\
{}& - \frac{\mu^2}{\sin^2\theta}\,\left(\frac{\hat{\mathbb{A}}}{\hat{\mathbb{B}}}\right)^{\delta^2-1}\,\psi + \Omega^2\, r^2\,\frac{{\hat{\mathbb{A}}^{\delta^2-2\,\delta}}}{{\hat{\mathbb{B}}^{\delta^2-1}}}\,\psi = 0\,.
\end{align}
Inspection of this wave equation reveals that it is separable in the very special case that $\mu = 0$ and $\Omega = 0$ regardless of the value of $\delta$; however, it is separable in general only for $\delta^2 = 1$, namely, in Schwarzschild spacetime. 

It is convenient to introduce the radial ``tortoise" coordinate $\xi$ appropriate for the exterior ($r > 2\,m$) region under consideration, namely, 
\begin{equation}\label{U6}
\xi = r + 2\,m \ln{\left(\frac{r}{2\,m} - 1 \right)}\,, \qquad    \frac{dr}{d\xi} = 1-\frac{2\,m}{r}\,, \qquad \psi(r, \theta) = \chi(\xi, \theta)\,.
\end{equation}
Then, 
\begin{equation}\label{U7}
r\,\frac{\partial}{\partial r}\left[\left(1-\frac{2\,m}{r}\right) r^2\,\frac{\partial}{\partial r}\left(\frac{\psi}{r}\right)\right]  = \frac{r^2}{1-\frac{2\,m}{r}} \left[\frac{\partial^2\chi}{\partial \xi^2} -\frac{2\,m}{r^3}\,\left(1-\frac{2\,m}{r}\right)\,\chi\right]\,
\end{equation}
and Eq.~\eqref{U5} takes the form
\begin{align}\label{U8}
{}& \frac{\partial^2\chi}{\partial \xi^2} + \Omega^2\,\frac{{\hat{\mathbb{A}}^{(\delta- 1)^2}}}{{\hat{\mathbb{B}}^{\delta^2-1}}}\,\chi -\frac{2\,m}{r^3}\,\left(1-\frac{2\,m}{r}\right)\,\chi \nonumber    \\
{}& + \frac{1}{r^2}\,\left(1-\frac{2\,m}{r}\right)\,\left[\frac{1}{\sin\theta}\,\frac{\partial}{\partial \theta}\left(\sin\theta\,\frac{\partial \chi}{\partial \theta}\right) - \frac{\mu^2}{\sin^2\theta}\,\left(\frac{\hat{\mathbb{A}}}{\hat{\mathbb{B}}}\right)^{\delta^2-1}\,\chi\right] = 0\,.
\end{align}

For $\delta = 1$, $\delta$-spacetime reduces to the Schwarzschild spacetime; then, as is well known, we can write $\chi = R_j (\xi)\,P_{j}^{\mu}(\theta)$, where $P_{j}^{\mu}$ is an associated Legendre polynomial with $j = 0, 1, 2, \cdots$, and $-j \le \mu \le j$, such that~\cite{A+S} 
\begin{equation}\label{U9}
\frac{1}{\sin\theta}\,\frac{d}{d \theta}\left(\sin\theta\,\frac{d P_{j}^{\mu}}{d \theta}\right) - \frac{\mu^2}{\sin^2\theta}\,P_{j}^{\mu}  = -j (j+1)\,P_{j}^{\mu}\,
\end{equation}
and the scalar wave equation reduces to the radial equation
\begin{equation}\label{U10}
\frac{d^2R_j}{d\xi^2}+\left(\Omega^2-V \right)R_j = 0\,, \qquad V=\left(1-\frac{2\,m}{r}\right) \left[\frac{j(j+1)}{r^2}+\frac{2\,m}{r^3} \right]\,,
\end{equation}
from which the standard theory of QNMs for scalar waves can be deduced. However, for $\delta = 1+q$, $q \ne 0$, we see that Eq.~\eqref{U8} is not separable and we have to resort to other methods. 

Imagine the wave amplitude that would correspond in the ray picture to the decay of the lightlike rings in the equatorial plane of a Schwarzschild black hole. Equation~\eqref{U9} for the associated Legendre polynomials implies that 
\begin{equation}\label{U11}
\left(\frac{1}{P_{j}^{\mu}}\,\frac{d^2P_{j}^{\mu}}{d \theta^2}\right)_{\theta = \pi/2} = \mu^2 -j (j+1)\,.
\end{equation}
In particular, for the light rings with $\mu = \pm j$ and $j \gg 1$, we note that the scalar wave amplitude corresponds to a wave packet concentrated around the equatorial plane with a small thickness of order $\pi/\sqrt{j}$ radians. Indeed, we have~\cite{A+S}
\begin{equation}\label{U11a}
P_{j}^{j} = (-1)^j \frac{(2j)!}{2^j\,j!}\, (\sin\theta)^j\,, \qquad P_{j}^{-j} = \frac{1}{2^j\,j!}\,(\sin\theta)^j\,,
\end{equation}
where the concentration of $(\sin\theta)^j$ about the equatorial plane ($\theta = \pi/2$) is illustrated in Figure 3 for $j \gg1$.
This feature should be essentially independent of the oblateness of the Schwarzschild source for small $q$, $0 < q \ll1$. Therefore, we henceforth assume that for $\mu = \pm j$ and $j \gg 1$,
\begin{equation}\label{U12}
\left[\frac{1}{\chi \,\sin\theta}\,\frac{\partial}{\partial \theta}\left(\sin\theta\,\frac{\partial \chi}{\partial \theta}\right)\right]_{\theta = \pi/2}  = \mu^2 -j (j+1) = -j\,,
\end{equation}
just as would be the case in Eq.~\eqref{U11}.

\begin{figure}
\includegraphics[scale=0.6]{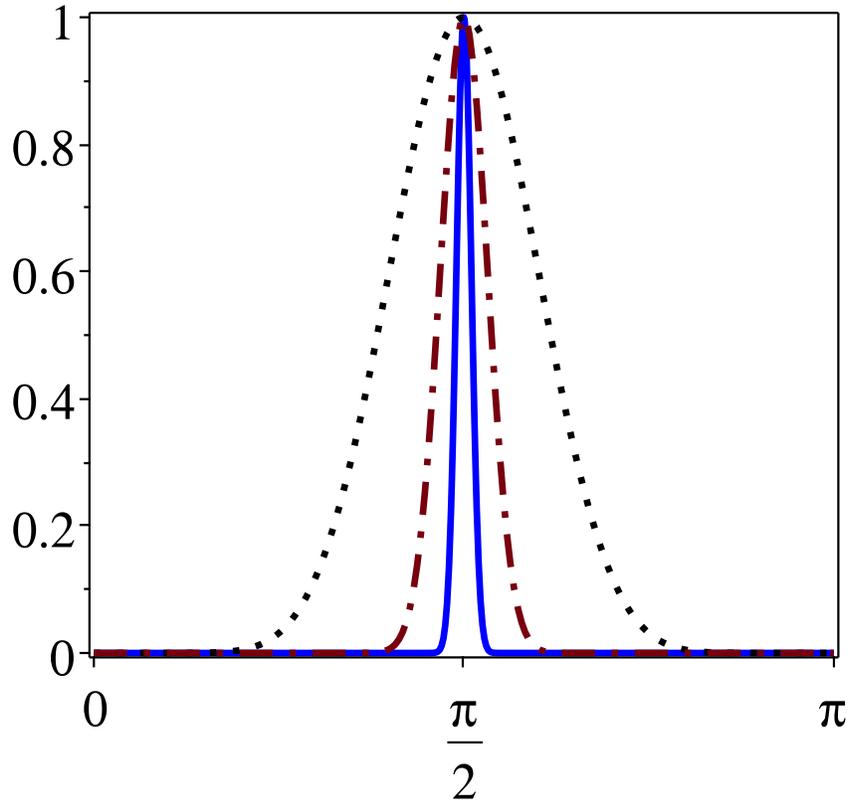}
\caption{\label{fig3} 
Plot of $(\sin\theta)^j$ versus $\theta$ for $j = 10$ (black dot), $j = 100$ (brown dot-dash) and $j = 1000$ (solid blue). The full width of this function at half maximum decreases as $j^{-1/2}$ with increasing $j$.}
\end{figure}

For $j \gg 1$, the wave packet that in the wave description corresponds to the decay of the light rings is essentially confined to the equatorial plane ($\theta = \pi/2$). The main wave dynamics contained in the scalar wave Eq.~\eqref{U8} thus occurs in the equatorial plane. In this connection, we define a new radial amplitude
\begin{equation}\label{U13}
 W(\xi) := \chi(\xi, \pi/2)\,
\end{equation} 
and with $\hat{\mathbb{B}} = \hat{\mathbb{A}} + m^2/r^2$, we can write Eq.~\eqref{U8} in the form
\begin{equation}\label{U14}
\frac{d^2 W}{d\xi^2} + \Omega^2\,\frac{{\hat{\mathbb{A}}^{(\delta- 1)^2}}}{{\hat{\mathbb{B}}^{\delta^2-1}}}\,W 
 - \frac{1}{r^2}\,\left(1-\frac{2\,m}{r}\right)\,\left[\frac{2\,m}{r} + j + j^2\,\left(\frac{\hat{\mathbb{A}}}{\hat{\mathbb{B}}}\right)^{\delta^2-1}\right] W = 0\,,
\end{equation}
where we have used Eq.~\eqref{U12} in accordance with our approximation scheme ($\mu = \pm j$, $j \gg 1$ and $\theta = \pi/2$). Here, $\delta = 1 + q$ and we work only to linear order in $q$; therefore, in Eq.~\eqref{U14}, $(\delta-1)^2 = q^2$  and $\delta^2- 1 = 2q + q^2$, where $q^2$ terms will be neglected. Thus, using the relation $X^q \approx 1+q\,\ln{X}$ for $X > 0$, Eq.~\eqref{U14} can be written as
\begin{equation}\label{U15}
\frac{d^2 W}{d\xi^2} +\left[\Omega^2 - \left(\frac{j(j+1)}{r^2}+\frac{2\,m}{r^3} \right)\,\hat{\mathbb{A}} -2\,q\left(\Omega^2 \,\ln{\hat{\mathbb{B}}} + \frac{j^2}{r^2}\,\hat{\mathbb{A}}\,\ln{\frac{\hat{\mathbb{A}}}{\hat{\mathbb{B}}}}\right)\right]\,W = 0\,, 
 \end{equation}
which for $q=0$ reduces to the radial equation in the Schwarzschild case~\eqref{U10}. Finally, we note that in the exterior region of $\delta$-spacetime, $r > 2\,m$; therefore, $j^2 \gg 1 > (2\,m)/r$, $j(j+1)$ can be replaced by $j^2$ and Eq.~\eqref{U15} simplifies to 
\begin{equation}\label{U16}
\frac{d^2 W}{d\xi^2} +\left[\Omega^2 - \frac{j^2}{r^2}\,\hat{\mathbb{A}} -2\,q\left(\Omega^2 \,\ln{\hat{\mathbb{B}}} + \frac{j^2}{r^2}\,\hat{\mathbb{A}}\,\ln{\frac{\hat{\mathbb{A}}}{\hat{\mathbb{B}}}}\right)\right]\,W = 0\,. 
 \end{equation}

\subsection{Radial Equation}

The gravitational source of $\delta$-spacetime is a deformed Schwarzschild black hole of mass $M = \delta\,m$. We want to view the radial Eq.~\eqref{U16} in terms of the scalar perturbation of the exterior of a slightly deformed Schwarzschild black hole. To this end, we formally replace $r$ by $r / \delta$, $\xi$ by $\xi /\delta$ and $W$ by $W /\delta^2$ in Eq.~\eqref{U16} to get
\begin{equation}\label{U17}
\frac{d^2 W}{d\xi^2} +\left[\Omega^2 - \mathcal{V}(\xi; \Omega)\right]\,W = 0\,, 
\end{equation}
where 
\begin{equation}\label{U18}
\xi = r + 2\,M\,\ln{\left(\frac{r}{2\,M} -1\right)}\, 
\end{equation} 
and
\begin{equation}\label{U19}
\mathcal{V}(\xi; \Omega) =  \frac{j^2}{r^2}\,\left(1-\frac{2\,M}{r}\right) + 2\,q\,\left[\Omega^2\,(1+\ln{\tilde{\mathbb{B}}}) + \frac{j^2}{r^2}\,\tilde{\mathbb{A}}\,\ln{\frac{\tilde{\mathbb{A}}}{\tilde{\mathbb{B}}}}\right]\,, 
\end{equation} 
where $\tilde{\mathbb{A}} = 1 -2\,M/r$ and $\tilde{\mathbb{B}} = \tilde{\mathbb{A}} + M^2/r^2$; see Eq.~\eqref{W11}. In this case, the potential $\mathcal{V}$ depends explicitly upon the frequency $\Omega$ and the QNM boundary conditions are satisfied when $q$ is such that $0 < q < (\sqrt{5} -1)/2$~\cite{Allahyari:2018cmg}. 

The problem of finding the QNM frequencies when the potential depends upon the frequency has been treated before~\cite{Mashhoon:1985cya} using the first-order WKB approximation, which is equivalent to matching the potential to an inverted harmonic oscillator potential~\cite{Ma, LiuMa}. Equivalently, the QNM frequencies can be obtained from the bound states of the inverted potential via a certain transformation~\cite{Mashhoon:1985cya}. In fact, for $q = 0$,  the corresponding  QNM frequencies for $j \gg 1$ are given by~\cite{Ma}
\begin{equation}\label{U20}
\Omega_{Sch}  = \frac{1}{3\sqrt{3}M}\,\left[ j +  i\,\left(n + \frac{1}{2}\right)\right]\,,\qquad n = 0, 1, 2, 3, \cdots\,
\end{equation}
and we proceed to approximate $\mathcal{V}$ with the potential of an inverted harmonic oscillator.
 
Let $\mathcal{V}_{max}$ be the value of the potential  $\mathcal{V}$ at its maximum, which occurs for $\xi_{max}$, $\mathcal{V}_{max} = \mathcal{V} (\xi_{max}; \Omega)$. The potential near its maximum can be approximated by an inverted harmonic oscillator potential, namely, 
\begin{equation}\label{U21}
\mathcal{V}  =  \mathcal{V}_{max} - \frac{1}{2} \kappa \,(\xi - \xi_{max})^2\,,
\end{equation}  
where $\kappa$ is the curvature of the potential at its maximum, 
\begin{equation}\label{U22}
\kappa  = -\left(\frac{d^2 \mathcal{V}}{d\xi^2}\right)_{\xi_{max}}\,.
\end{equation}   
Then, according to the first-order WKB approximation~\cite{Ma, LiuMa}, the \emph{complex} QNM frequency $\Omega$ is given by 
\begin{equation}\label{U23}
\Omega^2  =  \mathcal{V}_{max} + i\, \sqrt{2\,\kappa}\, \left(n + \frac{1}{2}\right)\,,\qquad n = 0, 1, 2, 3, \cdots\,.
\end{equation}   
 
The maximum of the potential $\mathcal{V}$ occurs at $r_{max}$, corresponding to $\xi_{max}$, where $d\mathcal{V}/dr = 0$. A simple calculation using  Eq.~\eqref{U19} results in
\begin{equation}\label{U24}
r_{max} = 3M + q\,M (1+ \mathcal{X})\,, \qquad \mathcal{X} := 27\,\frac{M^2 \,\Omega^2}{j^2}\,
\end{equation}  
and 
\begin{equation}\label{U25}
\mathcal{V}_{max} = \frac{j^2}{27\,M^2}\,\left\{1 + 2\, q\,\left[\ln{\frac{3}{4}} + \mathcal{X} \left( 1 + \ln{\frac{4}{9}}\right)\right]\right\}\,.
\end{equation}   
Furthermore, after a somewhat long but straightforward calculation, we find 
\begin{equation}\label{U26}
\sqrt{2\,\kappa} = \frac{2\,j}{27\,M^2}\,\left[1 + \frac{1}{4} \,q\,\left(1-\mathcal{X} + 4\, \ln{\frac{3}{4}}\right)\right]\,.
\end{equation}    
Substituting Eqs.~\eqref{U25} and~\eqref{U26} in Eq.~\eqref{U23}, the complex value of $\mathcal{X}$, proportional to the square of the QNM frequency $\Omega$, can be easily determined and is given by
\begin{equation}\label{U27}
\mathcal{X} = 1 + 2\,q (1-\ln{3}) + i\, \frac{2}{j}\,\left(n + \frac{1}{2}\right)\left[1 + q\,\left(2- \ln{3} +  \ln{\frac{4}{9}}\right)\right] + O \left(\frac{1}{j^2}\right)\,,
\end{equation}    
where  $n = 0, 1, 2, 3, \cdots$. Using the definition of $\mathcal{X}$ in Eq.~\eqref{U24}, namely, $\mathcal{X} = 27\,M^2 \,\Omega^2/j^2$, we find
\begin{equation}\label{U28}
\Omega  = \frac{1}{3\sqrt{3}M}\,\left\{ j\,[ 1 + q\,( 1-\ln{3})] + i\left(n + \frac{1}{2}\right)\left[1 +q\,\left(1 + \ln{\frac{4}{9}}\right)\right]\right\}\,,\quad n = 0, 1, 2, 3, \cdots\,,
\end{equation}
which coincides with $\Omega_{SQ} $ given by Eq.~\eqref{U0}.  Therefore, according to the complementary wave picture, the QNM frequencies of the $\delta$-spacetime agree with those calculated via the ray picture using the light-ring method within the framework of our eikonal approximation scheme.

\section{DISCUSSION}

In our recent detailed paper~\cite{Allahyari:2018cmg} as well as the present work, we have introduced, within the standard framework of Einstein's theory of gravitation, the notion of \emph{generalized black holes} and have computed some of their QNM frequencies.  In this way, one can now investigate gravitational waves emitted by deformed black holes involving small nonrelativistic  quadrupole and higher moments.  

A particular generalized black hole, namely, $\delta$- Kerr spacetime is discussed in detail in the present paper and its QNM frequencies are compared with those of the Hartle-Thorne spacetime discussed in our previous paper~\cite{Allahyari:2018cmg}. Furthermore, the scalar wave equation on $\delta$-spacetime is investigated for the first time and a novel analytic method is introduced to find the QNM frequencies even when the scalar wave equation is not separable. We demonstrate that the first-order WKB results in this case are identical to those obtained via the light-ring method that employs unstable null geodesics. 

More specifically, in the ray description, we approximate the $\delta$-Kerr metric by the Kerr-$Q$ metric which is valid to first order in the quadrupole moment $Q$ and  second order in the angular momentum $J$.  We have calculated certain QNM frequencies of the Kerr-$Q$ metric in the eikonal limit using the light-ring method.  Furthermore, in connection with the complementary wave description of the QNMs of the $\delta$-metric, we have presented a novel method for the calculation of the QNM frequencies of the massless scalar field perturbations of the exterior $\delta$-metric to linear order in $q = \delta -1$ using the first-order WKB approximation. The results of the wave picture are in agreement with those of the ray picture. 

In our previous work~\cite{Allahyari:2018cmg}, we calculated using the light-ring method the QNM frequencies of the rotating Hartle-Thorne metric for the same mass $M$, quadrupole moment $Q$ and angular momentum $J$ as in the Kerr-$Q$ metric. We find the remarkable result that the characteristic damped oscillations are almost essentially the same in these two different cases. It appears that for the same multipole moments, collapsed configurations respond  in the eikonal limit to exterior linear perturbations at late times in essentially the same way. This may be related to the fact that the multipole moments uniquely determine a stationary asymptotically flat vacuum spacetime.

\appendix

\section{Ray Description of the QNMs of Kerr-$Q$ Spacetime}

The black hole response to linear perturbations is generally dominated at late times by certain damped QNM oscillations that are directly related to the intrinsic properties of the black hole. To calculate the complex QNM frequencies, one may thus study the complete temporal evolution of any appropriate perturbation. To this end, we study the unwinding of unstable null circular equatorial geodesic orbits in Kerr-$Q$ spacetime.

\subsection{Light ring for Kerr-$Q$ spacetime}

We investigate the motion of rays in the equatorial plane of the Kerr-$Q$ spacetime. If $x^\mu(\ell)$ is a null geodesic with an affine parameter $\ell$, then
\begin{equation}\label{A1}
g_{\mu \nu}\, \frac{dx^\mu}{d\ell}\, \frac{dx^\nu}{d\ell} = 0\,,\qquad \frac{d^2x^\alpha}{d\ell^2} + \Gamma^\alpha_{\mu \nu}\, \frac{dx^\mu}{d\ell}\, \frac{dx^\nu}{d\ell} = 0\,.
\end{equation}
The Christoffel symbols are given  in Appendix B. We look for circular orbits with $r=r_0$ and $\theta=\pi/2$. In the Schwarzschild spacetime we have $r_{Sch}=3M$. The presence of rotation as well as the quadrupole moment modifies this relation. For a null path we find,
\begin{align}\label{A2}
g_{tt}+2g_{t\phi}\frac{d\phi}{dt}+g_{\phi\phi}\left(\frac{d\phi}{dt} \right)^2=0\,,
\end{align}
which, to first order in $q$ and second order in $a$, implies
\begin{align}\label{A3}
\left(\frac{d\phi}{dt}-\frac{2 a M}{r^3}\right)^2 = \frac{\tilde{\mathbb{A}}}{r^2}\,\left[1+2\,q\,\left(\frac{M}{r\tilde{\mathbb{A}}} +\ln{\tilde{\mathbb{A}}}\right) - \frac{a^2}{r^4\,\tilde{\mathbb{A}}}(r^2-8M^2)\right]\,.
\end{align}
Next, the radial component of the geodesic equation can be written as
\begin{align}\label{A4}
\Gamma^{r}_{tt}+2\Gamma^{r}_{t\phi}\frac{d \phi}{dt}+\Gamma^{r}_{\phi\phi}\left(\frac{d \phi}{dt} \right)^2 =0\,,
\end{align}
which implies
\begin{align}\label{A5}
\left(\frac{d\phi}{dt} + \frac{a M}{r^3}\right)^2 = \frac{M}{r^3}\,\left[1+q\,\left(\frac{M}{r\tilde{\mathbb{A}}} +2\,\ln{\tilde{\mathbb{A}}}\right) + \frac{2\,Ma^2}{r^3}\right]\,.
\end{align}
Equating Eqs.~\eqref{A3} and~\eqref{A5}, we find
\begin{align}\label{A6}
r_0&=3M-Mq\mp\frac{2 a}{\sqrt{3}}-\frac{2 a^2}{9M}\,,\\
\label{A7}   \frac{d\phi}{dt}&= \Omega_{\pm} = \pm\frac{1}{3\sqrt{3}M}\,\left[ 1 - q\, \left( -1+\ln{3}\right)  \pm \frac{2\,a}{3\sqrt{3}\,M} +\frac{11}{54}\, \frac{a^2}{M^2}\right]\,.
\end{align}

\subsection{QNM Frequencies}

Let us consider a slight perturbation of the null equatorial circular orbits. The perturbed orbits are given by
\begin{align}\label{A8}
r = r_0 [ 1 + \epsilon\,f(t)]\,, \qquad \phi = \Omega_{\pm} [t + \epsilon\,g(t)]\,, \qquad \ell = t + \epsilon\,h(t)\,.
\end{align}
The perturbed orbits must satisfy the null geodesic equations with the initial conditions
\begin{align}\label{A9}
f(0) = g(0) = h(0) = 0\,.
\end{align}
From Eq.~\eqref{A8} and the fact that the perturbed equatorial orbits are null rays, we have
\begin{align}\label{A10}
g_{tt}\left(\frac{dt}{d \ell} \right)^2+ g_{rr}\left(\frac{dr}{d \ell} \right)^2 +g_{\phi\phi}\left(\frac{d\phi}{d \ell} \right)^2+2\,g_{t\phi}\,\frac{dt}{d \ell}\frac{d\phi}{d \ell}=0 \,,\quad \frac{d\phi}{dt}= \Omega_{\pm}\,(1 + \epsilon\,g')\,,
\end{align}
where $g' := dg/dt$, etc.  After some algebra, we obtain $g'(t) = 0$; therefore, $g$ is a constant that must vanish in accordance with the initial conditions.

The time component of geodesic equation for $\theta = \pi/2$ leads to
\begin{align}\label{A11}
\frac{d^2t}{d\ell^2}+2\Gamma^{t}_{tr}\frac{dt}{d\ell}\frac{dr}{d\ell}+2\Gamma^{t}_{r\phi}\frac{dr}{d\ell}\frac{d\phi}{d\ell}=0\,.
\end{align}
Because $dr/d\ell \approx \epsilon\, r_0\,f'$, we need the Christoffel symbols in Eq.~\eqref{A11} to zeroth order in $\epsilon$. After some algebra, we find
\begin{align}\label{A12}
h''(t) = 2\, \left[1-q \pm\frac{a}{\sqrt{3}\,M} +\frac{4\,a^2}{9\,M^2}\right] f'(t)\,.
\end{align}
The radial component of the equatorial geodesic equation is 
\begin{align}\label{A13}
\frac{d^2r}{d\ell^2}+\Gamma^{r}_{tt}\left(\frac{dt}{d\ell} \right)^2+\Gamma^{r}_{\phi\phi} \left(\frac{d\phi}{d\ell} \right)^2 +2\Gamma^{r}_{t\phi}\frac{dt}{d\ell}\frac{d\phi}{d\ell}=0\,,
\end{align}
where we have omitted $\Gamma^{r}_{rr}$ term because it is second order in $\epsilon$. The radial contribution gives 
\begin{align}\label{A14}
 f''(t) - \frac{1}{27M^2}\,\left\{1 +2\,q\,[1 + 2\ln(2/3)]-\frac{4a^2}{27M^2}\right\} f(t) = 0\,.
\end{align}
For $\theta = \pi/2$, the geodesic equation for the azimuthal component is
\begin{align}\label{A15}
\frac{d^2\phi}{d\ell^2}+2\Gamma^{\phi} _{tr}\frac{dt}{d\ell}\frac{dr}{d\ell}+2\Gamma^{\phi}_{r\phi}\frac{dr}{d\ell}\frac{d\phi}{d\ell}=0\,
\end{align} 
and it yields the same result as Eq.~\eqref{A12}.

The  solutions for Eqs.~\eqref{A12} and~\eqref{A14} consistent with the initial conditions for $f$ and $h$ are
\begin{align}\label{A16}
f(t)&= \sinh(\zeta_{KQ}t), \\
\label{A17} h(t)&= 2\,\zeta_{KQ}^{-1}\left[1-q \pm \frac{a}{\sqrt{3}\,M} +\frac{4\,a^2}{9\,M^2}\right]\,\left[\cosh(\zeta_{KQ}t)-1 \right]\,, 
\end{align}
where
\begin{align}\label{A18}
\zeta_{KQ} =\frac{1}{3\sqrt{3}M}\,\left\{1 + q\,[1 + 2\ln(2/3)]-\frac{2a^2}{27M^2}\right\}\,.
\end{align}
Using the method described in detail in Section V of Ref.~\cite{Allahyari:2018cmg}, we can relate the imaginary part of quasinormal mode frequencies to $\zeta_{KQ}$, namely,  
\begin{align}\label{A19}
\Gamma_{KQ}  = \zeta_{KQ} \left( n + \frac{1}{2} \right)\,, \qquad n = 0, 1, 2, 3, \cdots\,.
\end{align}
Alternatively, one can use the method of Lyapunov exponents~\cite{Cardoso:2008bp}. Thus, we find the QNM frequencies that are given in Eq.~\eqref{W12}.

\section{ Christoffel symbols for Kerr-$Q$ metric}

The nonzero components of the Christoffel symbols of the Kerr-$Q$ metric~\eqref{W10} are given by
\begin{equation}\label{B1}
\Gamma^{t}_{tr}=\frac{M}{r^2\,\tilde{\mathbb{A}}}\,\left(1-q\,\frac{2 M}{r\,\tilde{\mathbb{A}}}\right)+a^2\frac{M}{r^5\tilde{\mathbb{A}}^2}\,\left[\left(4M-3r \right)\cos^2\theta-2M\sin^2\theta \right]\,, 
\end{equation}
\begin{equation}\label{B2}   
\Gamma^{t}_{r\phi}=-\frac{3aM}{r^2\tilde{\mathbb{A}}}\sin^2\theta,\qquad    \Gamma^{t}_{t\theta}=-\frac{a^2M}{r^3}\sin2\theta\,, 
\end{equation}
\begin{align}\label{B3}
\Gamma^{r}_{tt}&=\frac{M\tilde{\mathbb{A}}}{r^2}\,\left[1+ 2q\,\left(\frac{M}{r\,\tilde{\mathbb{A}}}+\ln\tilde{\mathbb{B}}\right)\right]+a^2\frac{M}{r^4}\left(1-4\tilde{\mathbb{A}}\cos^2\theta \right)\,, 
\end{align}
\begin{align}\label{B4}
\Gamma^{r}_{rr} ={}&-\frac{M}{r^5\,\tilde{\mathbb{A}}^2\,\tilde{\mathbb{B}}}\,\left\{r^3\,\tilde{\mathbb{A}}\,\tilde{\mathbb{B}}-2q\,M\,[r^2\,\tilde{\mathbb{A}}+ (r^2- 3\,Mr+3\,M^2)\,\sin^2\theta\,]{}\right\}\nonumber\\
{}&+\frac{a^2}{r^5\tilde{\mathbb{A}}^2}\,\left(r^2 -M\,r - r^2\,\tilde{\mathbb{A}}^2\,\cos^2\theta  \right)\,,
\end{align}
\begin{equation}\label{B5}
\Gamma^{r}_{r\theta} = \Gamma^{\theta}_{\theta \theta }=-\left(\frac{q\,M^2}{r^2\tilde{\mathbb{B}}} +\frac{a^2}{2r^2}\right)\,\sin2\theta\,,  \qquad  \Gamma^{r}_{t\phi} =-a\frac{M\tilde{\mathbb{A}}}{r^2}\sin^2\theta\,,
\end{equation}
\begin{align}\label{B6}
\Gamma^{r}_{\theta \theta} =-\frac{1}{r^2\,\tilde{\mathbb{B}}}\,\left[r^3\,\tilde{\mathbb{A}}\,\tilde{\mathbb{B}}+q\,M(r^2 - 2 Mr \cos^2\theta -M^2\sin^2\theta){}\right]+\frac{a^2}{r}\,\left( -1+\tilde{\mathbb{A}}\,\cos^2\theta\right)\, ,  
\end{align} 
\begin{align}\label{B7}
\Gamma^{r}_{\phi \phi }= {}&- r\,\tilde{\mathbb{A}}\,\left[1+q\,\left(\frac{M}{r\,\tilde{\mathbb{A}}} + 2\,\ln\frac{\tilde{\mathbb{B}}}{\tilde{\mathbb{A}}}\right)\right]\,\sin^2\theta \nonumber   \\
{}&+\frac{a^2}{r^3}\,\left[Mr\left(1-3\cos^2\theta \right)- (r^2+2M^2)\sin^2\theta  \right]\sin^2\theta\,,
\end{align}
\begin{align}\label{B8}
\Gamma^{\theta }_{rr}&=\left(\frac{q\,M^2}{r^4\tilde{\mathbb{A}}\tilde{\mathbb{B}}}+\frac{a^2}{2r^4\tilde{\mathbb{A}}}\right)\sin2\theta\,, \quad \Gamma^{\theta}_{\phi t}=a\frac{M}{r^3}\sin2\theta\,,\quad \Gamma^{\theta}_{tt} =-a^2\frac{M}{r^5}\sin2\theta\,,  
\end{align}
\begin{equation}\label{B9}
\Gamma^{\theta }_{r\theta}=\frac{1}{r^4\,\tilde{\mathbb{A}}\,\tilde{\mathbb{B}}}\,\{r^3\,\tilde{\mathbb{A}}\,\tilde{\mathbb{B}}-q\,M\,[r^2-2Mr(1+\sin^2\theta) + 3 M^2 \sin^2\theta\,]{}\}-\frac{a^2}{r^3}\cos^2\theta\,,     
\end{equation}
\begin{equation}\label{B10}
\Gamma^{\theta }_{\phi  \phi }=-\frac{1}{2}\left(1 + 2q\,\ln\frac{\tilde{\mathbb{B}}}{\tilde{\mathbb{A}}}\right)\,\sin 2\theta-\frac{a^2}{r^3}\left(4M+r \right) \sin^3\theta\cos\theta\,,  
\end{equation} 
\begin{equation}\label{B11}
\Gamma^{\phi }_{r\phi }=\frac{1}{r}\left(1-q\,\frac{M}{r\,\tilde{\mathbb{A}}}\right)+\frac{a^2}{2r^4\tilde{\mathbb{A}}}\left( -2r+M+3M\cos2\theta\right)\,,
\end{equation}
\begin{equation}\label{B12}
\Gamma^{\phi}_{tr}=\frac{aM}{r^4 \tilde{\mathbb{A}}}\,, \qquad \Gamma^{\phi}_{\theta t}=-\frac{2 a M}{r^3}\cot\theta\,,\qquad \Gamma^{\phi }_{\theta \phi }=\cot\theta+\frac{a^2 M}{r^3}\sin2\theta\,.
\end{equation}


\end{document}